\documentclass[prb,aps,twocolumn,showpacs,preprintnumbers,amsmath,amssymb,groupedaddress,superscriptaddress,letterpaper,floatfix]{revtex4-1}

\usepackage{amsmath}
\usepackage{graphicx}
\usepackage{dcolumn}
\usepackage{bm}

\begin{document}

\title{Proximity induced pseudogap in mesoscopic superconductor/normal-metal bilayers}

\author{Guo-Qiao Zha}%
\affiliation{Departement Fysica, Universiteit Antwerpen,
Groenenborgerlaan 171, B-2020 Antwerpen, Belgium}
\affiliation{Department of Physics, Shanghai University, Shanghai
200444, China}
\author{Lucian Covaci}%
\affiliation{Departement Fysica, Universiteit Antwerpen,
Groenenborgerlaan 171, B-2020 Antwerpen, Belgium}
\author{Shi-Ping Zhou}%
\affiliation{Departement Fysica, Universiteit Antwerpen,
Groenenborgerlaan 171, B-2020 Antwerpen, Belgium}
\affiliation{Department of Physics, Shanghai University, Shanghai
200444, China}
\author{F. M. Peeters}%
\affiliation{Departement Fysica, Universiteit Antwerpen,
Groenenborgerlaan 171, B-2020 Antwerpen, Belgium}

\begin{abstract}
Recent scanning tunneling microscopy measurements of the proximity
effect in Au/La$_{2-x}$Sr$_{x}$CuO$_{4}$ and
La$_{1.55}$Sr$_{0.45}$CuO$_{4}$/La$_{2-x}$Sr$_{x}$CuO$_{4}$ bilayers
showed a proximity-induced pseudogap [Yuli $et$ $al$., Phys. Rev.
Lett. {\bf 103}, 197003 (2009)]. We describe the proximity effect
in mesoscopic superconductor/normal-metal bilayers by using the Bogoliubov-de Gennes equations
for a tight-binding Hamiltonian with competing antiferromagnetic and
$d$-wave superconductivity orders . The temperature dependent local
density of states is calculated as a function of the distance from the interface. Bound state due to both d-wave and spin density wave gaps are formed in the normal metal for energies less than the respective gaps. If there is a mismatch between the Fermi velocities in the two layers we observe that these states will shift in energy when spin density wave order is present, thus inducing a minigap at finite energy. We conclude that the STM measurement in the proximity structures is able to distinguish between the two scenarios proposed for the pseudogap (competing or precursor to superconductivity). 
\end{abstract}

\pacs{74.45+.c, 74.50.+r}

\maketitle

One of the most important issues in cuprate high temperature
superconductors is the existence of a pseudogap (PG) phase in the
underdoped region \cite{TB,SM}. It has received heavy debate about
its nature as its origin may hold key information regarding the
pairing mechanism in these materials. One paradigm assumes that the
PG is only a manifestation of a competing or coexisting order of
superconductivity and it has no direct relation to the pairing
\cite{SC,CMV,JL}. However, other scenarios regard the PG as a
superconducting precursor state \cite{YJ,SA,ZY,JC,YL,HHW}, with close relation to Cooper pairing.
Despite many studies on the subject, it remains unclear
whether the PG phenomenon is related to superconductivity or not.

Very recently, the temperature evolution of the proximity-induced
gap in Au/La$_{2-x}$Sr$_{x}$CuO$_{4}$ and
La$_{1.55}$Sr$_{0.45}$CuO$_{4}$/La$_{2-x}$Sr$_{x}$CuO$_{4}$ bilayers
was reported by Yuli $et$ $al.$ \cite{OI} using scanning tunneling
microscopy (STM) and spectroscopy (STS) measurement to examine
whether unique spectral properties associated with pairing are
present above the superconducting transition temperature $T_{c}$ in
the PG temperature regime. For bilayers comprising an underdoped
La$_{2-x}$Sr$_{x}$CuO$_{4}$ (LSCO) film, there exists a smooth
evolution of the superconductor proximity gap into a
proximity-induced PG, as the temperature was raised above $T_{c}$.
In contrast, the proximity gap disappeared close to $T_{c}$ in
bilayers comprising an overdoped LSCO layer. They claimed that the
similar spatial dependence of the proximity induced PG and the
superconductor proximity gap indicates that the origin of the PG is
related to superconductivity. How to understand these behaviors
theoretically is still a challenging question. The proximity effect
between a superconductor and a normal metal has been thoroughly
studied using various techniques \cite{JH,JZ,JX,KH,LC}. However, an
investigation of the proximity effect in normal-metal/superconductor
bilayers above $T_{c}$ is still lacking.

In the present paper we use numerical solutions
of the Bogoliubov-de Gennes (BdG) equations based on a model
Hamiltonian with competing $d$-wave superconducting (DSC) and
antiferromagnetic (AFM) orders at finite temperature and
examine the proximity-induced gap in mesoscopic
normal-metal/superconductor bilayers. In cuprate superconductors,
the spin density wave (SDW) order with stripe modulation and the
accompanying charge order may emerge at high temperature $T$, and
could exist above the DSC transition temperature $T_{c}$
\cite{HYC}. Our numerical analysis focuses on the interplay between
DSC and SDW orders when in proximity with the metal as well as the
local density of states (LDOS) as a function of temperature and distance from the
interface.  States induced in the normal metal region depend on both DSC and SDW orders present in the superconducting side and on the Fermi velocity mismatch at the interface. Our results provide important new insight into the interpretation of recent STM experiments \cite{OI}.

In order to describe the superconductor/normal-metal bilayers we use
the tight-binding extended Hubbard Hamiltonian by assuming that the
on-site repulsion $U_{\mathbf{i}}$ is responsible for the competing
AFM order and the nearest-neighbor attraction $V_{\mathbf{ij}}$ for
the DSC pairing in the superconducting region:
\begin{eqnarray}
\label{eq:hamilt}
H&=&-\sum_{\langle\mathbf{ij}\rangle,\sigma}t_{\mathbf{ij}}c_{\mathbf{i}\sigma}^{\dagger}c_{\mathbf{j}\sigma}+\sum_{\mathbf{i},\sigma}(U_{\mathbf{i}}\langle{n_{\mathbf{i}\bar{\sigma}}}\rangle-\mu)c_{\mathbf{i}\sigma}^{\dagger}c_{\mathbf{i}\sigma}
\nonumber
\\&&+\sum_{\langle\mathbf{ij}\rangle}(\Delta_{\mathbf{ij}}c_{\mathbf{i}\uparrow}^{\dagger}c_{\mathbf{j}\downarrow}^{\dagger}+\Delta_{\mathbf{ij}}^{\ast}c_{\mathbf{j}\downarrow}c_{\mathbf{i}\uparrow}),
\end{eqnarray}
where $t_{\mathbf{ij}}=t$ are the nearest-neighbor hopping integral.
$c_{\mathbf{i}\sigma}$ ($c_{\mathbf{i}\sigma}^{\dagger}$) are
destruction (creation) operators for electron of spin $\sigma$,
$n_{\mathbf{i}\sigma}=c_{\mathbf{i}\sigma}^{\dagger}c_{\mathbf{i}\sigma}$
is the number operator, and $\mu$ is the chemical potential
determining the averaged electron density $\bar{n}$. The SDW and DSC
orders have the following definitions respectively:
$\Delta_{\mathbf{i}}^{SDW}=U_{\mathbf{i}}\langle{c_{\mathbf{i}\uparrow}^{\dagger}c_{\mathbf{i}\uparrow}-c_{\mathbf{i}\downarrow}^{\dagger}c_{\mathbf{i}\downarrow}}\rangle$
and
$\Delta_{\mathbf{ij}}=V_{\mathbf{ij}}\langle{c_{\mathbf{i}\uparrow}c_{\mathbf{j}\downarrow}-c_{\mathbf{i}\downarrow}c_{\mathbf{j}\uparrow}}\rangle/2$.
Using the Bogoliubov transformation,
$c_{\mathbf{i}\sigma}=\sum_{n}[u_{\mathbf{i}\sigma}^{n}\gamma_{n\sigma}-{\sigma}v_{\mathbf{i}\sigma}^{n\ast}\gamma_{n\bar{\sigma}}^{\dag}]$,
the Hamiltonian in Eq. (1) can be diagonalized by solving the
resulting BdG equations self-consistently,
\begin{equation}
\sum_{\mathbf{j}}^{N}\left(
               \begin{array}{cc}
                 \mathcal {H}_{\mathbf{ij}\sigma} & \Delta_{\mathbf{ij}} \\
                 \Delta_{\mathbf{ij}}^{\ast} & -\mathcal {H}_{\mathbf{ij}\bar{\sigma}}^{\ast} \\
               \end{array}
             \right
)\left(
   \begin{array}{c}
     u_{\mathbf{j}\sigma}^{n} \\
     v_{\mathbf{j}\bar{\sigma}}^{n} \\
   \end{array}
 \right)=E_{n}\left(
   \begin{array}{c}
     u_{\mathbf{i}\sigma}^{n} \\
     v_{\mathbf{i}\bar{\sigma}}^{n} \\
   \end{array}
 \right),
\end{equation}
where $\mathcal
{H}_{\mathbf{ij}\sigma}=-t_{\mathbf{ij}}+[U_{\mathbf{i}}\langle{n_{\mathbf{i}\bar{\sigma}}}\rangle-\mu]\delta_{\mathbf{ij}}$
. Using open boundary conditions we obtain the eigenvalues
{$\{E_{n}\}$} with eigenvectors $\{\mathrm{u}^{n},\mathrm{v}^{n}\}$.

\begin{center}
\begin{figure}[t]
\includegraphics[angle=-90,width=\columnwidth]{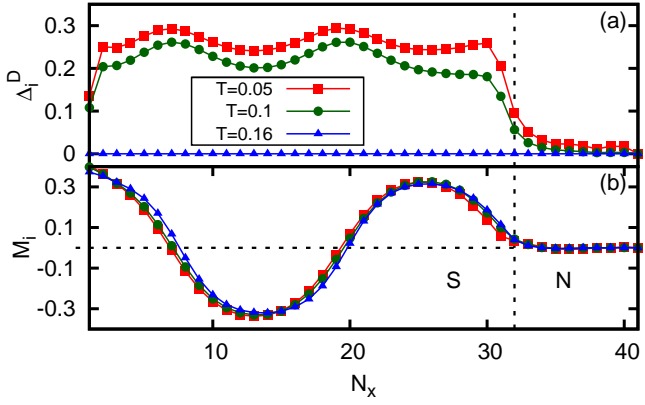}
\caption{ The spatial variation of DSC order parameter
$\Delta_{i}^{D}$ (a) and staggered magnetization $M_{i}^{s}$ (b) as
a function of $N_{x}$ for different temperatures $T$ in a mesoscopic
normal-metal/superconductor bilayer. ``S" and ``N" denote respectively the
superconducting region and normal metal. The 
parameter values are: $U=2.5$, $\bar{n}=0.9$, and $t^{'}=0$.}
\label{fig.1}
\end{figure}
\end{center}

The self-consistent conditions are:
\begin{gather}
\langle{n_{\mathbf{i}\uparrow}}\rangle=\sum_{n=1}^{2N}|\mathbf{u}_{\mathbf{i}}^{n}|^{2}f(E_{n}), \\
\langle{n_{\mathbf{i}\downarrow}}\rangle=\sum_{n=1}^{2N}|\mathbf{v}_{\mathbf{i}}^{n}|^{2}[1-f(E_{n})], \\
\Delta_{\mathbf{ij}}=\sum_{n=1}^{2N}\frac{V_{\mathbf{ij}}}{4}(\mathbf{u}_{\mathbf{i}}^{n}\mathbf{v}_{\mathbf{j}}^{n\ast}+\mathbf{v}_{\mathbf{i}}^{n\ast}\mathbf{u}_{\mathbf{j}}^{n})\tanh(\frac{{E_{n}}}{2k_{B}T}),
\end{gather}
where $f(E)=(e^{{E}/k_{B}T}+1)^{-1}$ is the Fermi-Dirac distribution
function. The DSC order parameter is defined at site $i$ as
$\Delta_{\mathbf{i}}^{D}=(\Delta_{\mathbf{i}+\mathbf{e}_{\mathbf{x}},\mathbf{i}}+\Delta_{\mathbf{i}-\mathbf{e_{x}},\mathbf{i}}-\Delta_{\mathbf{i},\mathbf{i}+\mathbf{e_{y}}}-\Delta_{\mathbf{i},\mathbf{i}-\mathbf{e_{y}}})/4$,
where $\mathbf{e_{x,(y)}}$ denotes the unit vector along the $x,(y)$
direction. Throughout this work, the distance is measured in units
of the lattice constant $a$, and the energy is scaled to $t$. In the
numerical calculations, we take $k_{B}=a=t=1$ for simplicity. The
averaged electron density is chosen as $\bar{n}$=0.9, corresponding
to the underdoped level $\delta$=0.1. For an appropriate initial
order parameter profile, the Hamiltonian is numerically diagonalized
and the obtained electron wave functions are used to calculate the
new parameters for the next iteration step. The solution is found
when the relative error in the gap function between successive
iterations is less than the desired accuracy.

\begin{center}
\begin{figure}[t]
\includegraphics[angle=-90,width=\columnwidth]{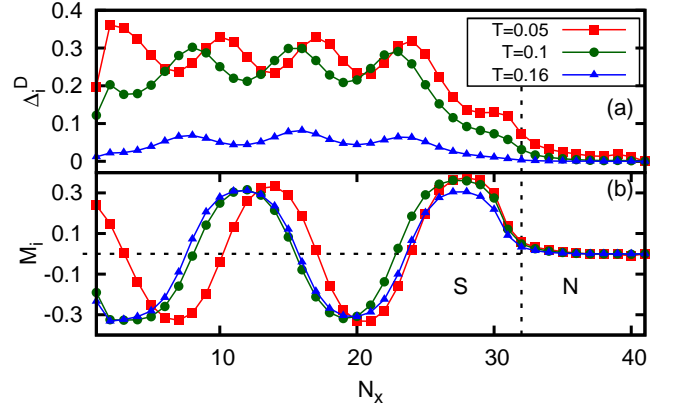}
\caption{ The spatial variation of $\Delta_{i}^{D}$ (a)
and $M_{i}^{s}$ (b) as a function of $N_{x}$ for different
temperatures $T$ in a mesoscopic normal-metal/superconductor
bilayers. The parameter values are: $U=2.5$, $\bar{n}=0.9$,
and $t^{'}=-0.2$.}
\label{fig.2}
\end{figure}
\end{center}

The LDOS of the energy $E$ at the position $\mathbf{i}$ can be
written as
\begin{equation}
\rho_{\mathbf{i}}(E)=\sum_{n=1}^{2N}[|\mathbf{u}_{\mathbf{i}}^{n}|^{2}\delta(E-E_{n})+|\mathbf{v}_{\mathbf{i}}^{n}|^{2}\delta(E+E_{n})],
\end{equation}
where $\delta(E)$ is the Dirac delta-function, which is broadened into
${\frac{1}{\pi}\frac{\Gamma}{(E-E')^{2}+\Gamma^{2}}}$,
with $\Gamma=0.01$ being chosen. $\rho_{\mathbf{i}}(E)$ is
proportional to the local differential tunneling conductance which
could be measured by STM experiments.

First, we consider mesoscopic superconductor / normal-metal bilayers
with sizes of $N_{x}$$\times$$N_{y}$=32$\times$32 (12$\times$32)
for the superconducting (normal) region. The
spatial variation of the DSC order parameter $\Delta_{i}^{D}$ (a) and
the staggered magnetization $M_{i}^{s}$ (b) along the $x$ axis for
various temperatures $T$ are plotted in Fig.~\ref{fig.1} for chosen parameter
values: $U=2.5t$ and $V=1.2t$ and $t'=0$. Here, the staggered magnetization of the induced AFM or SDW order is defined as
$M_{i}^{s}=(-1)^{i}(n_{i\uparrow}-n_{i\downarrow})$. Fig.~\ref{fig.2} shows the DSC and SDW order parameters for the same pairing potentials but for a different next-nearest neighbor hopping $t'=-0.2t$ in order to properly account for the band structure of high-$T_c$ cuprates. One
can easily see that, for both cases, the $y$-oriented stripe
modulations of DSC and coexisting SDW orders are present at lower temperatures. Note that, the oscillation periods decrease with
increasing the next-nearest neighbor hopping strength, while the period of the SDW is almost half of that of the DSC. Moreover, the periodicity of the
strip-modulated DSC and SDW orders for $t'=0$ seem not to be sensitive to
$T$. When the temperature increases, the DSC order is suppressed and finally disappears at $T_{c}$. Further increasing $T$, the staggered
magnetization decreases to zero at the Ne$\acute{\mathrm{e}}$l
temperature $T_{N}=0.3$. In this case the nature of the pseudogap is the competing SDW order present only at low doping and higher temperature. This coexistence could be the reason for the modulation of the superconducting gap observed in STM experiments on underdoped cuprates \cite{MV}.

\begin{center}
\begin{figure}[t]
\includegraphics[angle=-90,width=\columnwidth]{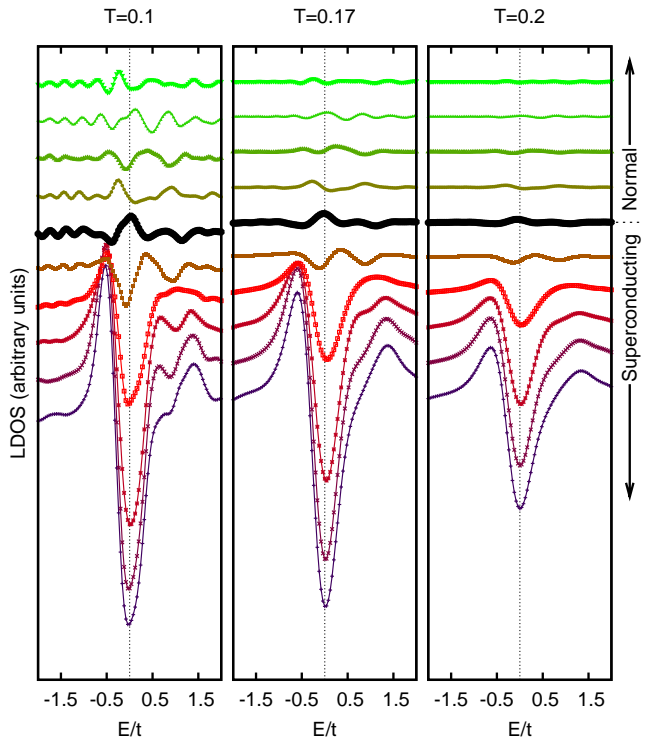}
\caption{ The LDOS for positions around the interface
for $t^{'}=-0.2t$ with $T=0.1$, $T_{c}=0.17$ and $T=0.2$.
The other parameter values are: $U=2.5$ and $\bar{n}=0.9$. The thick black line represents the LDOS at the interface while lower(upper) lines are in the SC(normal) region. The normal state LDOS is subtracted for all lines in order to suppress oscillations due to the interface with vacuum.}
\label{fig.3}
\end{figure}
\end{center}

Shown in Figs.~\ref{fig.1}~and~\ref{fig.2}, both DSC (panel (a)) and SDW (panel (b)) leak into the normal region due to the proximity effect. The usual exponential decay is observed, with a leaking distance which is of the order of the lattice constant. Similar to bilayers made of conventional SC and normal metals, bound states will be induced in the normal region for energies less than the superconducting gap. In order to see this we plot in Fig.~\ref{fig.3} the LDOS around the SC/normal metal interface. Similar to Ref.~\onlinecite{HYC} we find that in the superconducting region as the temperature is increased above $T_c$ a gap still exists due to the coexisting SDW order. Unfortunately, due to the discreteness of the energy spectrum and the finite temperature smearing, we cannot draw meaningful conclusions from the LDOS calculated in the normal-metal region. The main observation from Fig.~\ref{fig.3} is the modification of the LDOS in the normal metal region for energies below the DSC or SDW gaps.

In order to better understand the formation of bounds states in the normal region we performed a slightly different calculation. With the insight given by the self-consistent calculation done on a smaller sized system, we calculated the LDOS for a bilayer of size $N_x\times N_y=300\times 300$, with a normal metal region of size $L=4$. While still using the same model Hamiltonian we do not recalculate the order parameters self-consistently but consider them as step functions. If we are interested only in the LDOS at the surface of the normal-metal this approximation is reasonable since all the order parameters are set to zero in the normal region while if the coherence length is small it results in a sharp drop of the order parameters at the SC/N interface. Now instead of diagonalizing the full Hamiltonian we approximate the Green's functions and hence the LDOS in terms of Chebyshev polynomials. The procedure is presented elsewhere \cite{LC2}.

\vspace{-0.85cm}
Similar to an Andreev reflection at the N/S interface for quasi-particles with energies lower than the superconducting gap, at the N/AFM interface a peculiar reflection (Q-reflection \cite{BA,BI}) occurs if the energy is less than the SDW gap. Quasi-particles with momentum $\mathbf{k}_F$ are scattered to states with momentum $\mathbf{k}_F+\mathbf{Q}$ where $\mathbf{Q}$ is defined by the AFM order. If there is a potential barrier at the interface or a mismatch in the Fermi velocity, in addition to the Q-reflection, specular reflection also occurs thus influencing the formation of the bound states. 

\begin{center}
\begin{figure}[t]
\includegraphics[angle=-90,width=\columnwidth]{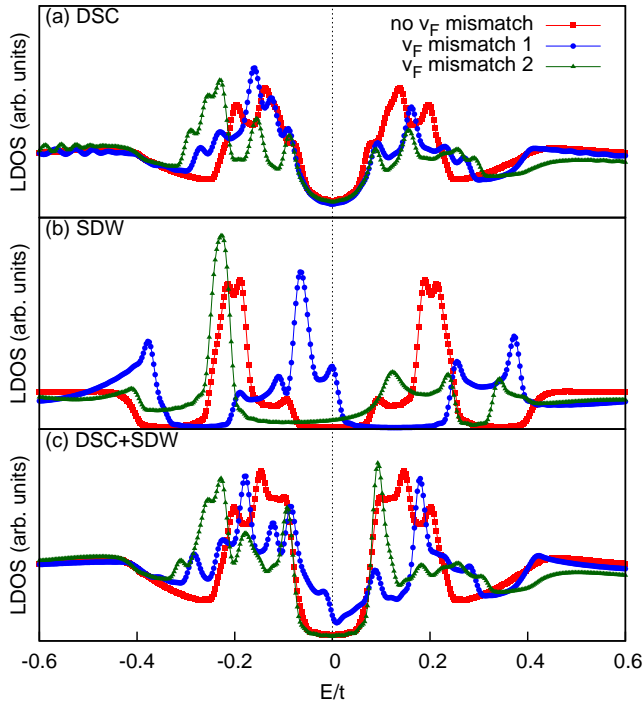}
\caption{ The LDOS at the surface of the normal metal region for three choices of order parameters: (a) $\Delta_d=0.4t$, $\Delta_{SDW}=0.0$,  (b) $\Delta_d=0.0$, $\Delta_{SDW}=0.4t$, (c) $\Delta_d=0.4t$, $\Delta_{SDW}=0.1t$. The lines represent three Fermi velocity mismatches at the N/SC interface: no mismatch, mismatch due to change in chemical potential and mismatch due to change in next-nearest neighbor hopping.}
\label{fig.4}
\end{figure} 
\end{center}

The main point in the STM measurement \cite{OI} was the fact that the observed gap is always at zero energy whether the normal metal layer is overdoped LSCO or Au. To simulate these conditions we consider three situations: no Fermi velocity mismatch, Fermi velocity mismatch from a shift in the chemical potential in the normal region ($\delta \mu=0.5$) or Fermi velocity mismatch from a different band structure ($t'=0$ in the SC region while $t'=-0.2t$ in the normal region). In Fig.~\ref{fig.4} we plot the calculated LDOS for three choices of the order parameters. First, in Fig.~\ref{fig.4}(a) only DSC ($\Delta_d=0.4t$) is considered. Subgap structures appear for all three choices of interfaces but the minigap is independent of this choice. Second, in Fig.~\ref{fig.4}(b) only SDW ($\Delta_{SDW}=0.4t$) is considered. Subgap features are also observed but depending whether there is Fermi velocity mismatch or not, the bound states shift in energy. The minigap now shifts to higher energy. Third, in Fig.~\ref{fig.4}(c) we consider a coexistence of DSC and SDW ($\Delta_d=0.4t$, $\Delta_{SDW}=0.1t$). In this case a minigap is still observed when the next-nearest neighbor hopping is changed in the normal layer although the particle-hole symmetry is broken. When the chemical potential is shifted in the normal metal, a suppression of the LDOS is still observed but the gap is closed. This behavior is not consistent to the experimental STM findings \cite{OI}, which showed that the minimum of the proximity induced pseudogap is always at the Fermi level irrespective of the normal metal type (overdoped $LSCO$ or $Au$). In order to explain the observed proximity induced pseudogap only scenarios which involve reflections with $\bf{Q}=0$, e.g. fluctuating SC order parameters or loop current orders \cite{VA}, should be considered.

In conclusion, we have investigated the proximity effect in
mesoscopic normal-metal/superconductor bilayers by numerically
solving the BdG equations based on an effective model Hamiltonian
with competing AFM and $d$-wave superconductivity interactions. The self-consistent solution gives oscillating DSC and SDW order parameters with the period of oscillations determined by the value of the next-nearest neighbor hopping amplitude. For low doping there exists a temperature range above $T_c$ in which only the SDW order is stable. As expected we observe in the metallic region the proximity induced pair correlations and the localized bound states for energies less that the DSC (or SDW) gap. More importantly by using a larger system we were able to confirm the conclusion presented in Ref.~\onlinecite{OI} which argues that an SDW order above $T_c$ would shift the minimum gap energy if the normal metal is changed from overdoped $LSCO$ to $Au$. By using a model microscopic Hamiltonian we showed that a mismatch in the Fermi velocity at the N/SC interface would shift the induced minigap. Depending on the nature of this mismatch (potential barrier or different band structure) the shift should also be observed below $T_c$ in the temperature range where DSC and SDW order coexist.

{\bf Acknowledgments}: This work was supported by the Flemish Science Foundation (FWO-Vl),
by Belgian Science Policy (IAP), by National Natural Science
Foundation of China under Grant No. 10904089 and No. 60971053, by
the Research Fund for the Doctoral Program of Higher Education of
China under Grant No. 20093108120005, by Shanghai leading academic
discipline project under Grant No. S30105, by Science and Technology
Committee of Shanghai Municipal under Grant No. 09JC1406000, by
Shanghai Municipal Education Committee under Grant No. shu-08053 and
No. 10zz63, and by Innovation Funds of Shanghai University.

\end{document}